\documentclass[12pt,draft]{article}
 
\newcommand{\be}{\begin{equation}}
\newcommand{\ee}{\end{equation}}
\newcommand{\bea}{\begin{eqnarray}}
\newcommand{\eea}{\end{eqnarray}}
\newcommand{\beaa}{\begin{eqnarray*}}
\newcommand{\eeaa}{\end{eqnarray*}}

\newcommand{\del}{\partial}
\newcommand{\g}{{\bf g}}
\newcommand{\h}{{\bf h}}
\newcommand{\ah}{{\hat \alpha}}
\newcommand{\bh}{{\hat \beta}}
\newcommand{\gh}{{\hat \gamma}}
\newcommand{\A}{{\cal A}}
\newcommand{\D}{{\cal D}}

\newcommand{\K}{{\bf k}}

\newcommand{\BB}{{{\rm I} \kern -2pt \rlap {\rm B} \kern +8pt}}

\newcommand{\half}{{\textstyle{1\over2}}}

\newcommand{\Pf}{{\rm Pf}}
\newcommand{\hl}{\\ \hline \\}

\advance\textheight by \topskip
\makeatletter

\@addtoreset{equation}{section}

\def\section{\@startsection {section}{1}{\z@}{-3.5ex plus -1ex minus
 -.2ex}{2.3ex plus .2ex}{\large\bf\centering}}
\def\subsection{\@startsection{subsection}{2}{\z@}{-3.25ex plus%
 -1ex minus -.2ex}{1.5ex plus .2ex}{\bf}}
\def\subsubsection{\@startsection{subsubsection}{3}{\z@}{-3.25ex plus%
 -1ex minus -.2ex}{1.5ex plus .2ex}{\sl}}
\makeatother

\begin{document}

\baselineskip 18pt
\parindent 12pt
\parskip 10pt

\begin{titlepage}
\begin{flushright}
PUPT-1922\\
DAMTP-2000-28\\
Imperial/TP/99-00/20\\
hep-th/0003264v2\\
March 2000\\[3mm]
\end{flushright}
\vspace{.4cm}
\begin{center}
{\Large {\bf
Commuting charges and symmetric spaces}}\\
\vspace{1cm}
{\large J.M. Evans${}^{a,b}$\footnote{e-mail: evans@feynman.princeton.edu, 
J.M.Evans@damtp.cam.ac.uk}, 
A.J. Mountain${}^c$\footnote{e-mail: A.Mountain@ic.ac.uk}}
\\
\vspace{3mm}
{\em ${}^a$ Joseph Henry Laboratories, 
Princeton University, Princeton NJ 08544, U.S.A.}\\
{\em ${}^b$ DAMTP, University of Cambridge, Silver Street, Cambridge
CB3 9EW, U.K.}\\
{\em ${}^c$ Blackett Laboratory, Imperial College, Prince Consort Road,
London SW7 2BZ, U.K.}\\
\end{center}

\vspace{1cm}
\begin{abstract}
\noindent
Every classical sigma-model with target space a compact 
symmetric space $G/H$ (with $G$ classical) 
is shown to possess infinitely many local, 
commuting, conserved charges which can be written in closed form.
The spins of these charges run over a characteristic
set of values, playing the role of exponents of $G/H$, and 
repeating modulo an integer $h$ which plays the role of a 
Coxeter number.

\end{abstract}

\end{titlepage}

\section{Introduction}
An important set of classically integrable field theories in 
1+1 dimensions is provided by non-linear sigma-models with target 
manifold a symmetric space $G/H$.
These have been studied for many years, particularly in 
connection with a one-parameter `dual symmetry' which allows the 
construction of both local and non-local conserved quantities.
Little is known about the local charges arising from this approach, 
however, with comparatively few examples which can be written 
in closed form. A comprehensive discussion, with many references, 
can be found in \cite{EF}. Examples of more recent work 
involving various models based on symmetric spaces are 
\cite{Fendley,Schwarz,Mira,Park}.

Principal chiral models (PCMs) can be regarded as special cases of 
symmetric space models (SSMs), with target manifold a Lie group $G$. 
In \cite{EHMM2}, it was shown that other well-known local 
conservation laws in the PCMs, apparently different from those arising 
from dual symmetry, exhibit striking and hitherto unexpected properties. 
In particular, it was possible to construct mutually-commuting sets of 
charges with a characteristic pattern of spins given by the exponents of 
each classical group $G$ modulo its Coxeter number. 
The same pattern of spins arises in affine Toda field theories, and proves 
central to understanding a number of their most important properties 
\cite{corri94,dorey91}. 

In this note we will show that results similar to those of \cite{EHMM2} 
can be obtained for any sigma-model based on a symmetric space $G/H$ 
with $G$ a classical group. Specifically, we will consider local conserved 
currents which can be written in a simple, closed form, and which lead to 
commuting sets of charges whose spins are related to the underlying 
symmetric space data.  
We begin by formulating the field theory and the relevant conservation laws.

Let $g(x^\mu)$ be a field on two-dimensional Minkowski space
taking values in some compact Lie group $G$, with Lie algebra $\g$.
Let $H \subset G$ be some subgroup and $\h \subset \g$ the corresponding 
Lie subalgebra. To formulate the sigma-model with target space 
$G/H$, we introduce a gauge field $A_\mu$ in $\h$ and define a 
covariant derivative
\be
D_\mu g = \del_\mu g - g A_\mu
\ee
with the property that 
\be
g \mapsto g h \, , \quad A_\mu \mapsto h^{-1} A_\mu h + h^{-1} \del_\mu h
\qquad \Rightarrow \qquad 
D_\mu g \mapsto (D_\mu g) h
\ee
for any function $h(x^\mu)$. It is also useful to introduce 
the $\g$-valued currents 
\be
j_\mu = g^{-1} \del_\mu g \, , \qquad 
J_\mu = g^{-1} D_\mu g  =  j_\mu - A_\mu \ .
\ee
The latter current is covariant under gauge transformations, with 
\be
J_\mu \mapsto h^{-1} J_\mu h \ .
\ee

The $G/H$ sigma-model is defined by the lagrangian
\be
{\cal L} = - {1\over 2} {\rm Tr} (J^\mu J_\mu) 
= - {1\over 2} {\rm Tr} (g^{-1} D^\mu g \, g^{-1} D_\mu g ) 
\ee
which has a global $G$ symmetry (acting from the left on $g$)
and the local $H$ symmetry discussed above.
The equation of motion for the field $g$ is 
\be\label{geqn}
D_\mu J^\mu = \del_\mu J^\mu + [ A_\mu , J^\mu ] = 0 \ .
\ee 
By combining this with the identity 
$\del_\mu j_\nu - \del_\nu j_\mu + [ j_\mu , j_\nu ] = 0$ 
we obtain
\be\label{lceqn}
2 \del_- J_+ + 2 [ A_- , J_+ ] = [J_+ , J_-]  + F_{+-}
\ee
where $F_{\mu \nu} = \del_\mu A_\nu - \del_\nu A_\mu + [ A_\mu , A_\nu ]$,
and we have made use of light-cone coordinates, $V_\pm = V_0 \pm V_1$.
The equations of motion for the $A_\mu$ fields are
\be\label{Aeqn} 
J_\mu = 0 \quad {\rm on}~~\h \ .
\ee

Everything we have said so far applies to an arbitrary homogeneous space
$G/H$. The special nature of symmetric spaces emerges when we look
for conserved quantities. It is helpful to recall what happens for the
PCM based on $G$, which can be obtained by specializing the analysis above 
to the case in which $H$ is trivial, setting $A_\mu = 0$, and hence 
$J_\mu = j_\mu$. The equation of motion then becomes
$\del_- j_+ = \half [ j_+ , j_-]$ which implies conservation
equations such as $\del_- {\rm Tr} (j_+^m) = 0$.
For a general, non-trivial gauge group $H$, the additional term $F_{+-}$
in (\ref{lceqn}) prevents one from carrying out a similar 
construction in any obvious way. 
But for $G/H$ a symmetric space there is an orthogonal decomposition of 
the Lie algebra, and a compatible ${\bf Z_2}$ grading of the Lie bracket,
\be\label{grade}
\g = \h + {\bf k} \, : \qquad 
[ \h , \h ] \subset \h
\, , \quad
[ \h , {\bf k} ] \subset {\bf k}
\, , \quad
[ {\bf k} , {\bf k} ] \subset \h \ .
\ee
Since the $A_\mu$ equations of motion force $J_\mu$ to take values in
${\bf k}$, this grading then implies that the left- and 
right-hand sides of (\ref{lceqn}) must vanish separately:
\be\label{lax}
\del_- J_+  = - [ A_- , J_+ ] \quad {\rm in}~{\bf k}
\, , \qquad
[J_+ , J_-] = - F_{+-} \quad {\rm in}~{\bf h} \ .
\ee
The first of these conditions allows the construction of conservation
laws very similar to those of the PCM,
{\it e.g.} $\del_- {\rm Tr} (J_+^m) = 0$.
A conserved current of this type can be 
written down using any symmetric invariant tensor, and we now discuss
the possibilities.

\section{Currents and invariants on $G/H$}

Let us introduce a basis of anti-hermitian generators $\{t^a\}$ for $\g$,
obeying $[ t^a , t^b ] = f^{abc} t^c$ and ${\rm Tr} (t^a t^b) = -\delta^{ab}$.
Since $\g$ is compact, we need not distinguish upper and lower Lie algebra 
indices, and the structure constants $f^{abc}$ are real and totally 
antisymmetric. We can assume our basis is chosen so that it splits 
$a \rightarrow (\ah , \alpha)$, with the subsets $\{t^\ah\}$ and 
$\{t^\alpha\}$ providing bases for $\h$ and $\K$ respectively in the 
decomposition in (\ref{grade}). The currents of the $G/H$ SSM model can 
now be written $j^a_\mu$, while the non-trivial components of the gauge 
fields are $A_\mu^\ah$. 
The symmetric space condition, or ${\bf Z}_2$ grading,
means that the non-vanishing structure constants are $f_{\ah \bh \gh}$, and 
$f_{\ah \beta \gamma }$, up to permutations of indices.

Consider some totally symmetric tensor $d^{(m)}_{a_1 \ldots a_m}$ 
of degree $m$ on $\g$ (we use `degree' rather than `rank' to avoid
confusion with the rank of the algebra; we shall not always indicate the 
degree explicitly). By virtue of its symmetry, this tensor is completely 
determined by the associated function 
$ d( X ) \equiv d_{a_1 \ldots a_m} X^{a_1} \ldots X^{a_m}$
where $X = X^a t^a$ is an arbitrary element of the Lie algebra.
If $\tau$ is any map from $\g$ to itself, then we define a new tensor 
$\tau(d)$ by $\tau (d) (X) = d( \tau(X) ) $.
We shall call $d$ a $G$-{\em invariant\/} tensor, or simply an 
{\em invariant tensor on\/} $\g$, if $d = \tau (d)$
whenever $\tau$ is an inner automorphism of $\g$,
so that $\tau (X) = gXg^{-1}$ for some $g$ in $G$.
This is equivalent to the condition 
\[
d^{(m)}_{ c (a_1 \ldots a_{m-1}} f^{\phantom{m}}_{a)b c} = 0 \ .
\]
Similar definitions apply in an obvious way to subgroups of $G$ acting 
on subspaces of $\g$.

A conserved charge of spin $s$ in the PCM based on $G$ can be constructed 
from each symmetric invariant tensor $d^{(s+1)}_{a_1 a_2 \ldots a_s a_{s+1}}$ 
\cite{EHMM2}. 
A related conservation law in the $G/H$ SSM arises by restricting such a 
tensor to $\K$, thus considering just the components 
$d^{(s+1)}_{\alpha_1 \alpha_2 \ldots \alpha_s \alpha_{s+1}}$. 
The invariance condition written above means that the restricted tensor obeys 
\be\label{hinv}
d^{(s+1)}_{\gamma (\alpha_1 \ldots \alpha_s} f_{\alpha) 
\hat \beta \gamma}
= 0 \ .
\ee
It is easy to check directly that this implies 
\be\label{cons}
\del_- ( d_{\alpha_1 \ldots \alpha_{s+1} } 
J_+^{\alpha_1} \ldots J_+^{\alpha_{s+1}} ) 
= 0
\ee
on using the equation of motion in (\ref{lax}). From this current
we obtain a conserved charge of spin $s$ in the usual way: 
\be\label{charge}
q_s = \int d x \, d_{\alpha_1 \ldots \alpha_{s+1}} 
J_+^{\alpha_1} \ldots J_+^{\alpha_{s+1}} \ .
\ee
Similar conserved charges can be constructed from $J_-^\alpha$; 
their properties are directly analogous and we will not discuss 
them further. 

It is important to check that the tensor $d^{(s+1)}$ does not vanish
when restricted to $\K$ in order to have a {\em non-trivial} conserved 
quantity in the SSM. We shall return to this point in section 6. 
It is also clear that (\ref{hinv}) and (\ref{cons}) rely only 
on the fact that the tensor is $H$-invariant on $\K$.
It is not obvious, a priori, that any such tensor should arise as the 
restriction of some $G$-invariant tensor on $\g$, but this emerges 
from the analysis below.
These and other matters can be understood in terms 
of the special role played by {\em primitive\/} invariants.

For each Lie algebra $\g$ there are exactly ${\rm rank}(G)$
primitive symmetric invariants, with the property that all others 
can be written as polynomial functions of them (see {\it e.g.}~\cite{azca97}).
This happens essentially because any element in $\g$ is conjugate 
to some element in a fixed Cartan subalgebra (CSA), and so any invariant
tensor is determined by its values on the ${\rm rank}(G)$ independent basis 
elements of this CSA.
The degrees of the primitive invariant tensors for each classical group 
$G$ are given in the table, in terms of the exponents, $s$. 

One convenient choice for the primitive invariants consists of symmetric 
traces, with $d(X) = {\rm Tr} (X^m)$ (of the appropriate degrees), 
together with the Pfaffian invariant
$d(X) = {\Pf} (X) \equiv \epsilon_{i_1 j_1 \ldots i_n j_n} 
X_{i_1 j_1} \ldots X_{i_n j_n}$ for the special case of $SO(2n)$.
Other choices are possible, and will be important later.
However, any choices of the primitive invariants differ 
only by terms which are products of polynomials of lower degrees.

We need to determine how these facts generalize from a group $G$ to a
symmetric space $G/H$. 
One can define a CSA for $G/H$ as a maximal set of 
mutually commuting generators in $\K$, and ${\rm rank} (G/H)$ is then 
the number of elements in such a set.
It can also be shown that any element of $\K$ is conjugate, by 
an element of $H$, to a member of some chosen CSA \cite{Helg}. 
This means that, just as for groups,
any invariant is determined by its values on the CSA, and there are 
precisely ${\rm rank}(G/H)$ primitive invariants, in terms of which all 
others can be expressed.

The degrees of the primitive invariants for each symmetric space
$G/H$ with $G$ classical are given by the data in the table.
Each primitive $H$-invariant tensor is obtained by restricting 
a primitive $G$-invariant tensor on $\g$ to the subspace $\K$. 
Other invariants which are primitive on $\g$ may vanish when 
restricted to $\K$, or else fail to be 
primitive on $\K$ in some more complicated fashion.
For our purposes we may take the values $s$ given in the table 
as a definition of the exponents of $G/H$.
The Pfaffian in $SO(2n)$ appears as something of a special case,
lying outside the regular sequence formed by the other invariants.
Because of this, we have chosen to separate the values of 
$s$ corresponding to the Pfaffian, or its restriction, by a semi-colon.

\[
  \begin{array}{cccc}
    G/H~\mbox{symmetric space} & \quad \mbox{rank} (G/H) \quad 
       & s : \, d^{(s+1)}~\mbox{primitive} & \quad h \quad \\ \hl
    SU(n) & n{-}1 & 1, 2, \dots , n{-}1 & n \\ \hl
    SO(2n{+}1) & n & 1, 3, \dots , 2n{-}1 & 2n\\ \hl
    SO(2n) & n & 1, 3, \dots , 2n{-}3;n{-}1 \quad & \quad 2n{-}2\\ \hl
    Sp(2n)  & n & 1, 3, \dots , 2n{-}1 & 2n \\ \hl
    {SU(p{+}q)}/S(U(p){\times}U(q)) \quad (p\leq q) & p 
      & 1, 3, \dots , 2p{-}1 & 2p \\ \hl
    {SO(p{+}q)}/{SO(p){\times}SO(q)} \quad (p < q) & p 
      & 1, 3, \dots , 2p{-}1 & 2p \\ 
    \\
    {SO(2n)}/{SO(n){\times}SO(n)} & n & 1, 3, \dots , 2n{-}3; n{-}1
      & 2n{-}2 \\ \hl
    {Sp(2p{+}2q)}/{Sp(2p){\times}Sp(2q)} \quad (p \leq q) 
    & p & 1, 3, \dots , 2p{-}1 & 2p
      \\ \hl
    SU(n)/SO(n) & n{-}1 & 1, 2, \dots , n{-}1 & n \\ \hl
    {Sp(2n)}/{U(n)} & n & 1, 3, \dots , 2n{-}1 & 2n  \\ \hl
    {SO(2n)}/{U(n)} & [n / 2] & 1, 3, \dots , 2 [ n / 2 ]{-}1 
     & 2 [n/2] \\ \hl
    {SU(2n)}/{Sp(2n)} & n{-}1 & 1, 2, \dots , n{-}1 & n \\ \hl
  \end{array}
\]

We obtained the results in the table by making use of convenient canonical 
forms for the CSA generators in $G/H$.
In view of the remarks above, it is the set of eigenvalues of the 
CSA generators, and how they behave under $H$, 
which determines the allowed invariants, and which of them 
are primitive. By comparing with the well-known data for classical groups, 
it is not difficult to arrive at the results for symmetric spaces. 
The method is best illustrated by some examples, but a fuller 
explanation of this sort requires some additional technical preparation,
and so we consign these details to a separate section below.
Our main task---understanding the conserved charges in the $G/H$ 
sigma-model---will then be resumed in section 4.

\section{Some details and examples}

Recall that the CSA of each classical 
Lie algebra can be parameterized by a set of real `eigenvalues'
$\lambda_i$ as follows
\begin{eqnarray}
su(n): &&{\rm diag} ( \, i\lambda_1 \, , \, \ldots \, , \, i\lambda_n \, )
\qquad {\rm with} \qquad \lambda_1 + \ldots + \lambda_n = 0
\nonumber\\[4pt]
so(2n): && {\rm diag} \left ( \, 
\left [ \matrix{0&\lambda_1\cr -\lambda_1&0\cr} \right ] , \, 
\ldots \, , \, 
\left [ \matrix{0&\lambda_n\cr - \lambda_n&0\cr} \right ] \, \right )
\nonumber\\[3pt]
so(2n{+}1): && {\rm diag} 
\left ( \, \left [ \matrix{ 0&\lambda_1 \cr -\lambda_1&0\cr} \right ] , 
\, \ldots \, , \, 
\left [ \matrix{0&\lambda_n\cr -\lambda_n&0\cr}\right ] , \, 0 \, \right )
\nonumber\\[3pt]
sp(2n): &&{\rm diag}
(\, i\lambda_1 \, , \, \ldots \, , \, i \lambda_n \, , \,  
- i \lambda_1 \, , \, \ldots \, , \, - i\lambda_n \, )
\nonumber
\end{eqnarray}
where we have used an obvious block notation for the orthogonal algebras.
The function $d(X)$ defined by a symmetric invariant tensor $d$ on 
$\g$ is some polynomial in the eigenvalues $\lambda_i$.
This polynomial must be totally symmetric,
because in all cases a member of the CSA of $\g$ can be conjugated 
by specific elements of $G$ so as to permute the eigenvalues in 
any desired way. 
(To show this it actually suffices to use the simple result
$ 
\pmatrix{ 0 & 1 \cr -1 & 0 \cr}  
\pmatrix{a & 0 \cr 0 & b \cr}
\pmatrix{ 0 & -1 \cr 1 & 0 \cr}  
= \pmatrix{b & 0 \cr 0& a \cr} 
$
in appropriate block forms.)

Now, for $\g = su(n)$ there are invariant tensors corresponding 
to all symmetric polynomials in the eigenvalues, 
except for $\lambda_1 + \ldots + \lambda_n= 0$.
The symmetrized traces mentioned earlier clearly give rise to the power 
sums $\sum_i \lambda_i^m$, and a basis of primitive invariants corresponds 
to the finite subset with $m=s{+}1$ and $s$ an exponent.
On the other hand, for $so(2n{+}1)$ and $sp(2n)$, only polynomials 
in {\em even\/} powers of $\lambda_i$ are allowed, since the sign of 
any eigenvalue can be reversed 
by conjugating with a suitable element of $G$. For example,
conjugating a CSA element of $\g = so(2n{+}1)$
with ${\rm diag} ( 1 , -1 , 0 , \ldots , 0 , -1)$ 
(which certainly belongs of $G = SO(2n{+}1)$) changes the sign of 
$\lambda_1$. 
For $\g = so(2n)$ it is also possible to reverse the signs 
of eigenvalues, but only in {\em pairs}. For instance, 
we can conjugate by ${\rm diag }( 1, -1 ,1 , -1 , 0 , \ldots , 0)$
which reverses the signs of $\lambda_1$ and $\lambda_2$, but we cannot 
conjugate by any element of $G= SO(2n)$ and change the sign of just one
eigenvalue. In addition to the even symmetric powers, these 
symmetry properties allow precisely one more independent invariant,
which is the Pfaffian, proportional to 
$\lambda_1 \ldots \lambda_n$.

In this manner the allowed invariants and primitive 
invariants for each Lie algebra are characterized as certain totally symmetric 
polynomials in the CSA eigenvalues. Moreover, we can distinguish three 
classes of polynomials, depending on their additional symmetry properties:
A-type, like $su(n)$---no additional symmetries; B/C-type,
like $so(2n{+}1)$ or $sp(2n)$---invariant under reversal of sign of each  
eigenvalue separately; D-type, like $so(2n)$---invariant under reversal 
of signs of pairs of eigenvalues.\footnote{
These symmetry operations on the CSA eigenvalues are 
usually presented as actions of the Weyl group.}

The results for each symmetric space $G/H$ 
can now be found by the following steps.
First, find a convenient parameterization of $\K$ and its CSA, as in
{\it e.g.}~\cite{Helg}; $H$-invariant tensors on $\K$ are 
polynomials in the `eigenvalues' of the CSA generators.
Next, examine the action (via conjugation) of specific elements of $H$ on 
these eigenvalues, and so determine that the polynomials 
have symmetry type A, B/C, or D using the terminology introduced above. 
Finally, check that all primitive invariants of this type indeed arise 
as restrictions of $G$-invariant tensors 
on $\g$, {\it e.g.}~by considering traces or symmetric 
powers. We now sketch how this works for some examples; the remaining
cases in the table can be handled similarly.
  
The Grassmannians $SO(p{+}q)/SO(p){\times}SO(q)$ and 
$SU(p{+}q)/S(U(p){\times}U(q))$ are conveniently treated 
together. For either family the subgroup $H \subset G$ has the block structure
$\pmatrix{ P & 0\cr 0 & Q \cr}$ and $\K$ consists of matrices 
$\pmatrix{ 0 & X \cr -X^\dagger & 0 \cr }$ 
where $P$ is $(p{\times}p)$, $Q$ is $(q{\times}q)$, 
and $X$ is $(p{\times}q)$ with real or complex entries. 
For both the real and complex families, the CSA can be parameterized by 
\[
X = \pmatrix{
\lambda_1 &0&\cdots&0&0&\cdots&0\cr 
0&\lambda_2&\cdots&0&0&\cdots&0 \cr              
\vdots&\vdots&\ddots&\vdots&\vdots&&\vdots\cr
0&0&\cdots&\lambda_p&0&\cdots&0\cr
}
\]
with real `eigenvalues' $\lambda_i$ (we assume $p\leq q$).
Notice that the effect on $\K$ of conjugating by a general element of $H$
is $X \mapsto P X Q^{-1}$. One can readily choose $P$ and $Q$ 
so as to permute the CSA eigenvalues in any desired way.
 
When $p < q$, it also easy to find elements of $H$ which change 
the sign of any given eigenvalue, just as in the previous discussion of 
$so(2n{+}1)$. This implies that the invariant polynomials  
are of B/C-type. They arise as restrictions of symmetric 
traces on $\g$, and in fact they can be written 
${\rm Tr} (XX^{\dag} \ldots XX^{\dag})$, which is manifestly 
invariant under $H$. 
When $p = q$, a new feature arises 
for the real Grassmannians: just as for the Lie algebra $so(2p)$, 
it is now only possible to change the signs of eigenvalues in pairs,
so the pattern of invariants is type D.
An additional primitive invariant on $\K$ arises as the restriction 
of the Pfaffian on $\g = so(2p)$, and it can be written
$\epsilon_{i_1 \ldots i_p} \epsilon_{j_1 \ldots j_p} X_{i_1 j_1} \ldots 
X_{i_p j_p}$. Note that under the action of $H$ this 
expression changes by a factor ${\rm det} (P) \, {\rm det} (Q^{-1})$
which is indeed unity for these examples. 
For the complex Grassmannians, however, the invariants are still of type B/C,
even when $p = q$. This is because there are elements in $H$ such as 
$P = Q^{-1} = {\rm diag} ( i, 1, \ldots, 1)$ which change the sign of
a single eigenvalue, in this case $\lambda_1$.
Consistent with this, such elements have 
${\rm det} (P) \, {\rm det} (Q^{-1}) \neq 1$.

The next example is 
$G/H = SU(n)/SO(n)$. The subspace $\h$ consists of 
real, antisymmetric matrices, while $\K$ consists of 
imaginary, symmetric, traceless matrices, and its natural 
CSA coincides with the standard choice for $su(n)$.
Invariance under $H = SO(n)$ means precisely that the 
eigenvalues can be permuted in any desired way.
The set of invariants is of type A, the same as those for $\g = su(n)$, 
and they clearly descend from these by restriction to $\K$.

Finally, consider $G/H = SU(2n)/Sp(2n)$. 
We can choose the subspaces $\h$ and $\K$ to consist of matrices 
with the block forms
$\pmatrix{A & B \cr -B^* & A^* \cr}$ and
$\pmatrix{C & D \cr D^* & -C^* \cr}$ respectively,
where $A$ lives in $u(n)$, $B$ is complex and symmetric, 
$C$ lives in $su(n)$, and $D$ is complex and antisymmetric.
The CSA elements are $X = \pmatrix{ Y & 0 \cr 0 & Y}$ 
where $Y$ is any generator in the standard CSA of $su(n)$. 
The set of primitive invariants is once again A-type, 
matching those of $su(n)$.
Unlike in the previous example, however, they arise here as 
restrictions of tensors on $\g = su(2n)$.

\section{Poisson brackets of currents and charges}

Returning now to our treatment of the $G/H$ sigma-model, 
we are interested in the classical Poisson brackets (PBs) of the currents 
$j^a_\mu$ and $J^a_\mu$. 
A convenient way to calculate these directly, without first finding 
PBs for the underlying fields $g$, is to use the approach of \cite{fadd87},
which is easily adapted to the present situation.
We take $j_1^a$ as an independent canonical coordinate and view 
$j^a_0$ as a non-local function of it by means of the identity 
$\del_0 j_1 - \del_1 j_0 + [ j_0 , j_1 ] = 0$, which implies 
\be\label{jzerodef}
j_0 = \D^{-1} (\del_0 j_1 )
\qquad {\rm where} \qquad 
\D X \equiv \del_1 X + [ j_1 , X ] \ .
\ee
(We must assume appropriate boundary conditions on the fields, to
allow the definition and manipulation of inverse differential operators.)
In addition, the field $A_1^\ah$ can be eliminated from the 
lagrangian immediately by its equation of motion, 
$J_1^\ah = j_1^\ah - A_1^\ah =0$.
On making these replacements, the lagrangian becomes 
\be 
{\cal L} = \half (\D^{-1} \del_0 j_1 - A_0)^a 
(\D^{-1} \del_0 j_1 - A_0)^a 
- \half j_1^\alpha j_1^\alpha \ .
\ee
The momentum conjugate to $j_1^a$ is
\be\label{pidef}
\pi^a = - (\D^{-2} \del_0 j_1 -  \D^{-1} A_0 )^a 
\ee
and we impose canonical, equal-time PBs
\be\label{canpb}
\{ j_1^a(x) , \pi^b (y) \} = \delta^{ab} \delta(x{-}y) \ .
\ee
The resulting Hamiltonian density is 
\be\label{ham} 
{\cal H} = \half J_0^a J_0^a + A_0^\ah J_0^\ah + \half j_1^\alpha j_1^\alpha
\ee
where
\be\label{Jzerodef}
J_0^a = -\D \pi^a \ .
\ee
The remaining, time-like component of the gauge field $A_0^\ah$ acts as a 
Lagrange multiplier, imposing the constraint
\be\label{const}
J_0^\ah \approx 0 \ .
\ee
(We have taken a well-known short-cut by applying Dirac's
procedure \cite{HT} without introducing momenta conjugate to the 
gauge fields, which play the role of Lagrange multipliers.)

To summarize: the independent canonical variables are $j_1^a$ and $\pi^a$,
obeying (\ref{canpb}). In terms of these, $j_0^a$ is defined by 
(\ref{jzerodef}); $J_1^\ah = 0$; $J_1^\alpha = j_1^\alpha$;
and $J_0^a$ is defined by (\ref{Jzerodef}).
The system is governed by the Hamiltonian density in (\ref{ham}),
together with the constraint (\ref{const}).
From (\ref{canpb}) it is straightforward to calculate
the equal time PBs 
\begin{eqnarray}
\left\{ J_0^a (x), J_0^b (y) \right\} 
& = & - f^{abc} \, J_0^c (x) \, \delta(x{-}y) \nonumber \\
\left\{ J_0^a (x), j_1^b (y) \right\} 
& = & - f^{abc} \, j_1^c (x) \, \delta(x{-}y) + \delta^{ab} 
\, \delta'(x{-}y) \label{PBs}\\
\left\{ j_1^a (x), j_1^b (y) \right\} 
& = & 0 \nonumber
\end{eqnarray}
which are the objects of central importance for us.
The first of these implies that the constraints (\ref{const})
are first-class, corresponding to the original $H$ gauge invariance.
(They are analogous to the Gauss Law constraint arising in
canonical treatments of electromagnetism.)
It is also a simple consequence of (\ref{PBs}) that the constraints 
weakly commute with the Hamiltonian, so they are preserved in time,
and the canonical formulation has therefore been completed in a 
consistent fashion.

Let us now consider the canonical brackets of two conserved charges
of type (\ref{charge}).
We need not concern ourselves with gauge fixing the remaining 
constraints (\ref{const}), because each current (\ref{cons}) commutes with 
them (each current is invariant under $H$-gauge transformations) and so the
Dirac bracket of two conserved charges, obtained after imposing
some gauge choice, is always identical to their Poisson bracket
\[
\{ q_s , q_r \} = 
\left \{ \, \int d x \, d^{(s+1)}_{\alpha_1 \ldots \alpha_{s+1}}
J_+^{\alpha_1} (x) \ldots J_+^{\alpha_{s+1}} (x)
\, , \,
\int d y \, d^{(r+1)}_{\beta_1 \ldots \beta_{r+1}}
J_+^{\beta_1} (y) \ldots J_+^{\beta_{r+1}} (y)
\, \right \}
\]
which can be calculated from (\ref{PBs}).
The terms in (\ref{PBs}) containing $\delta(x{-}y)$ do not 
contribute, by invariance of the $d$-tensors 
(the arguments are exactly similar to those given in \cite{EHMM2} 
for the PCM). The terms involving $\delta'(x{-}y)$, on the other 
hand, result in an integrand proportional to
\[
d^{(s+1)}_{\alpha_1 \ldots \alpha_s \gamma} \, 
d^{(r+1)}_{\beta_1 \ldots \beta_r \gamma} \,
J_+^{\alpha_1} \ldots J_+^{\alpha_{s}} J_+^{\beta_1} \ldots J_+^{\beta_{r-1}}
\del_1 J_+^{\beta_r} \ .
\]
The charges $q_s$ and $q_r$ will commute if and only if this 
integrand is a total derivative, which is true if and only if
\be\label{dident}
d^{(s+1)}_{(\alpha_1 \ldots \alpha_s}{}^\gamma 
d^{(r+1)}_{\beta_1 \ldots \beta_{r-1} ) \beta_r}{}^\gamma 
=
d^{(s+1)}_{(\alpha_1 \ldots \alpha_s}{}^\gamma
d^{(r+1)}_{\beta_1 \ldots \beta_r)}{}^\gamma \ .
\ee

It was shown in \cite{EHMM2} (see also eqn.~(2.39) of \cite{EHMM3}) 
that there exist tensors $k^{(s+1)}$ for each 
classical Lie group $G$ with the property that
\be\label{kident}
k^{(s+1)}_{(a_1 \ldots a_s}{}^c
k^{(r+1)}_{b_1 \ldots b_{r-1} ) b_r}{}^c
=
k^{(s+1)}_{(a_1 \ldots a_s}{}^c
k^{(r+1)}_{b_1 \ldots b_r)}{}^c \ .
\ee
This is exactly what is required to ensure commuting charges in the 
PCM based on $G$, and these tensors exist whenever $s$
is an exponent of $G$ modulo $h$. 
An obvious possibility is to choose the same tensors $k^{(s+1)}$ for $G/H$
as for $G$. We can certainly restrict the free indices
in (\ref{kident}) from $\g$ to the subspace $\K$,
but it is not clear that we can restrict the repeated 
index from $c$ in (\ref{kident}) to $\gamma$, so as to arrive 
at (\ref{dident}).
Such a restriction is allowed in all cases of interest, however, 
by virtue of the property
\be\label{vanish} 
d^{(s+1)}_{\alpha_1 \ldots \alpha_{s} \alpha_{s+1}} \neq 0 
\qquad \Rightarrow \qquad d^{(s+1)}_{\alpha_1 \ldots \alpha_s \gh} = 0 \ .
\ee 
We shall prove in the next section that this holds for any invariant 
tensor on a classical symmetric space.

\section{More properties of invariant tensors on $G/H$}

Let us return to the definition of $G/H$ via a ${\bf Z}_2$ 
grading of $\g$. An equivalent statement is that there is a Lie algebra 
automorphism $\sigma$ of $\g$ with $\sigma^2 =1$. The subspaces
$\h$ and $\K$ are the eigenspaces of $\sigma$ with 
eigenvalues $+1$ and $-1$ respectively.

An invariant tensor $d^{(m)}$ on $\g$ is not necessarily invariant under 
$\sigma$ (unless $\sigma$ is an inner automorphism of $\g$),
but let us assume for the moment that $\sigma (d) = d$.
Since elements of $\K$ change sign under $\sigma$, it follows that
if the degree $m$ is odd, then $d^{(m)}$ must vanish when restricted to $\K$.
If $m$ is even, $d^{(m)}$ need not vanish on $\K$, but 
then it must satisfy 
$d^{(m)}_{\alpha_1 \ldots \alpha_{m-1} \gh} = 0$, since $m{-}1$ is 
odd.

Now suppose that $d$ is not invariant under $\sigma$.
It turns out that in all such cases, $d$ is instead invariant under
the map $\tilde\sigma (X) \equiv - \sigma (X)$, as we shall see below.
The map $\tilde \sigma$ is not an automorphism of $\g$, and it has eigenspaces 
$\h$ and $\K$ with the reversed eigenvalues $-1$ and $+1$ respectively.
Requiring that $d^{(m)}$ be invariant under $\tilde \sigma$ and 
that it be non-zero on $\K$ does not result in 
any restriction on the degree $m$.
However, since $\tilde \sigma$ reverses the sign of each element in $\h$,
invariance of $d$ under this map does imply that 
$d^{(m)}_{\alpha_1 \ldots \alpha_{m-1} \gh} = 0$, as required.

We conclude that (\ref{vanish}) holds for every tensor 
$d$ obeying either $\sigma (d) = d$, or $\tilde \sigma (d) = d$.
To check that this exhausts all possibilities,
we can consider the simple explicit forms for $\sigma$ that are available
for each of the classical symmetric spaces \cite{Helg}.

For the three families of Grassmannian symmetric spaces
as well as for $SO(2n)/U(n)$ and $Sp(2n)/U(n)$, we can take 
$\sigma$ to be a map on $\g$ of the form 
\[
\sigma(X) = M X M^{-1} \ , 
\]
for some matrix $M$. 
(Once again, this is not necessarily an inner automorphism, 
because $M$ may not belong to $G$.) Clearly this implies 
$\sigma (d) = d$ whenever $d$ is a symmetrized trace.
The only primitive invariant not of this type is the 
(restriction of) the Pfaffian for $SO(2n)/SO(n){\times}SO(n)$.
In this case $\sigma (d) = (\det M) d = (-1)^n d$, since $M$ is 
diagonal with $n$ pairs of eigenvalues $\pm1$.
When $n$ is even, $\sigma(d) = d$, but when $n$ is odd, the degree 
of $d$ is also odd, and then $\tilde \sigma(d) = d$, as claimed.\footnote{
These remarks are consistent with the patterns of 
invariants given in earlier sections: some cosets always 
have $\sigma(d) = d$, and their invariants were already 
found to have even degrees.}

Finally we consider the families $SU(n)/SO(n)$ and $SU(2n)/Sp(2n)$ 
for which the automorphism of $\g$ can be written in the form
\[
\sigma (X) = M X^* M^{-1}  \qquad \Rightarrow \qquad 
\tilde \sigma (X) = M X^T M^{-1} \, , 
\]
since $X$ is anti-hermitian.
For the first family $M$ is the identity, while for the second 
it is some standard symplectic structure.
In either case it is clear that $\tilde \sigma (d) = d$ whenever
$d$ is a symmetrized trace, and it follows that the same is true 
for any tensor $d$. 

\section{The classical spectrum of commuting charges}

Our results concerning commuting charges in the $G/H$ SSM can now be 
summarized as follows. From each invariant tensor $k^{(s+1)}$ on 
$\g$ we obtain a conserved current in the $G/H$ model by restricting 
the tensor to $\K$, leading to a conserved charge of spin $s$.
These conserved charges always commute with one another, by virtue 
of the arguments given in sections 4 and 5 above.
The final issue we must settle is precisely which tensors $k^{(s+1)}$ 
are non-zero when restricted to $\K$, so as to give non-trivial
conserved charges for the $G/H$ SSM.

Let us first recall what happens for a group $G$ \cite{EHMM2}.
The tensors $k^{(s+1)}$ provide a set of primitive invariants 
when $s$ runs over the exponents of $G$.
These primitive invariants have $s < h$, the Coxeter number of $G$. 
For larger values of $s$, the tensors $k^{(s+1)}$ are not
primitive, but they are still non-zero (yielding non-trivial 
commuting charges) precisely when $s$ is equal to an exponent 
of $G$ modulo $h$.

We have discussed in detail in sections 3 and 4 the patterns
of invariants for each $G/H$, and how these are obtained from 
invariants on $G$. We can certainly choose our primitive tensors 
on $G/H$ from amongst the $k^{(s+1)}$ required for commuting charges. 
We have already defined the corresponding values of $s$ to be the exponents 
of the symmetric space $G/H$.
Similarly, we now define an integer $h$ for each $G/H$, with values 
specified in the table; this will play the role of the Coxeter 
number.\footnote{The discussion in section 3 effectively associates 
a Weyl group to each symmetric space, and our definition of $h$ 
coincides with the standard one for such groups
\cite{corri94}.}

With these definitions, our final result can be stated very simply:
$k^{(s+1)}$ is non-zero on $\K$ only when $s$ is equal to an exponent 
of $G/H$ modulo $h$. 
Thus, just as for groups, there are commuting charges associated with 
primitive invariants and hence exponents 
$s < h$, and also with non-primitive, but non-vanishing invariants 
for values of $s > h$ which repeat in families modulo $h$.

In some cases this result is automatic, because the values of 
$s$ run over all the positive odd integers, 
which trivially repeat modulo an even value of $h$. In other 
instances the pattern of spins is more involved, but the result for $G/H$ 
still follows directly from the corresponding statement for the group $G$. 
However, there is one family, namely $SU(2n)/Sp(2n)$, which 
requires special care.
For these spaces we have defined $h = n$, 
and so the validity of our result depends, in particular, upon the fact 
that the tensors $k^{(pn+1)}$ should vanish when restricted to $\K$ for 
any integer $p$. When $p$ is even we know that this tensor actually 
vanishes on $\g$, but when $p$ is odd it is non-zero on $\g$ and it is 
not immediately clear why it should vanish on $\K$.

To understand how this comes about we need to look more closely at the 
definition of the $k$-tensors for the algebras $su(n)$, and to make clear 
which algebra we are talking about we shall write $k^{(m)}_{su(n)}$.
Although these objects depend upon the value of $n$, they do so 
in a rather simple way, as revealed by the formula \cite{EHMM2}: 
\[
k^{(m)}_{su(n)}(X ) = \A^{(m)} \left (
\, {1 \over n} {\rm Tr} (X^m) ,
\, {1 \over n} {\rm Tr} (X^{m-1}) , 
\ldots , 
\, {1 \over n} {\rm Tr} (X^2) \, \right ) \ .
\]
The polynomials $\A^{(m)}$ are in fact characterized by the property
$k^{(n+1)}_{su(n)} = 0$, and it can be shown that 
this implies $k^{(pn+1)}_{su(n)} =0$ for any integer $p$ \cite{EHMM2}.

The coset $SU(2n)/Sp(2n)$ was one of the examples discussed earlier
in section 4. The CSA can be chosen to consist of 
matrices of the form $X = \pmatrix{ Y & 0 \cr 0 & Y \cr}$ where $Y$ is 
diagonal, traceless, and pure-imaginary, and hence belongs to the 
CSA of $su(n)$. Now clearly 
\begin{eqnarray*}
k^{(m)}_{su(2n)} (X) & = & \A^{(m)} \left (
\, {1 \over 2n} {\rm Tr} (X^{m}) ,
\, {1 \over 2n} {\rm Tr} (X^{m-1}),
\ldots , 
\, {1 \over 2n} {\rm Tr} (X^2) \, \right ) \\
& = &  \A^{(m)} \left (
\, {1 \over n} {\rm Tr} (Y^{m}) ,
\, {1 \over n} {\rm Tr} (Y^{m-1}),
\ldots , 
\, {1 \over n} {\rm Tr} (Y^2) \, \right ) \\
& = & k^{(m)}_{su(n)} (Y) \ .
\end{eqnarray*}
The last expression vanishes when $m = pn{+}1$ for any integer
$p$, as mentioned above. 
Since any invariant tensor is determined by its values on 
a CSA, we conclude that $k^{(pn{+}1)}_{su(2n)}$ indeed vanishes 
when restricted to $\K$, as claimed.

\section{Comments}

The results of this paper reinforce the elegant mathematical structure 
underlying the classical integrability of sigma-models on symmetric spaces. 
It would be interesting to investigate whether the charges we have 
constructed might be related to those arising in \cite{EF}, whose 
properties are rather more mysterious. There are also a number of other 
directions for future work. Although we have considered only classical 
groups, we expect similar results to hold for the exceptional groups and 
their symmetric spaces. At the quantum level, integrability is believed to 
depend upon $H$ being simple, and for these cases exact S-matrices have 
been proposed \cite{AAF}. It would be interesting to examine whether 
quantum-mechanical survival of our charges is consistent with these 
proposals, in light of the treatment of affine Toda theories in 
\cite{dorey91,corri94}.
Finally, one could extend the results of \cite{EHMM3} to supersymmetric 
$G/H$ models, with the added incentive that these are believed to be 
quantum-integrable for any symmetric space \cite{AF}.
\vskip 10pt

{\bf Acknowledgments}. We thank Jose Azc\'arraga and Tony Sudbery for 
helpful discussions. JME is supported by NSF grant PHY98-02484 and
by a PPARC Advanced Fellowship. AJM thanks the 
1851 Royal Commission for a Research Fellowship. 
This work was also supported in part by PPARC
under the SPG grant PPA/G/S/1998/00613.

\parskip=0pt

\end{document}